\documentclass[twocolumn,aps,amssymb,prb]{revtex4}
\usepackage{graphicx}
\usepackage{epsfig,amsmath}

\begin{document}

\title{Quantum transitions induced by the third cumulant of current fluctuations}

\author{T.~Ojanen}
\email[Correspondence to ]{teemuo@boojum.hut.fi}
\author{T. T. Heikkil\"a}
 \affiliation{ Low Temperature Laboratory, Helsinki University of
Technology, P.~O.~Box 2200, FIN-02015 HUT, Finland }

\date{\today}

\begin{abstract}
We investigate the transitions induced by non-Gaussian external
fluctuations on a small quantum system. The rates for the
transitions between the energy states are calculated using the
real-time Keldysh formalism for the density matrix evolution. We
detail the effects of the third cumulant of current fluctuations
coupled to a quantum system with a discrete level spectrum and
propose a setup for detecting the frequency-dependent third
cumulant through the transitions it induces. We especially discuss
a scheme where the fluctuations are coupled to a Josephson flux
qubit.
\end{abstract}
\pacs{PACS numbers: } \bigskip

\maketitle

The study of fluctuations has been in the center of interest in
physics for decades. The relevance of noise and fluctuations is
underlined by the fundamental relation between fluctuations and
dissipation in physical systems. One very concrete example of
fluctuations is the current noise in electric circuits. At
equilibrium, it obeys the fluctuation-dissipation theorem which
relates the magnitude of fluctuations to the temperature and the
impedance of the circuit. For a quantum system with a finite number
of levels interacting with an environment, the magnitude of these
fluctuations in the environment then determines the steady state of
the system, along with the rate with which this steady state is
approached.

During the past decade, the theory of electric fluctuations in
mesoscopic systems has been significantly developed to
characterize them also out of equilibrium, \cite{q,SZ} where a
finite average current leads to shot noise. The 
study yields information about the microscopic physical phenomena
inside electric conductors and the effects of the electromagnetic
environment on mesoscopic circuits. In large wires the current
statistics is Gaussian and fully characterized by the average
current and the noise power. The experimental development in
manufacturing smaller circuits has enabled the study of the
non-Gaussian character of fluctuations in mesoscopic samples.
\cite{ReuletBomze,lindell} In principle, the knowledge of these
fluctuations allows for an improved characterization of the
conductors, \cite{q} or the study of the effect their non-Gaussian
character causes on other mesoscopic systems.
\cite{tobiska,sonin,heikkila,ankerhold}

With a nonvanishing average current, the probability distribution
of current fluctuations no longer needs to be symmetric around the
average current. In particular, the third cumulant of fluctuations
describing the skewness of the current distribution may be finite.
It is also the lowest cumulant indicating a non-Gaussian
distribution. Despite the strong theoretical effort describing the
nature of the higher-order cumulants,\cite{q} measuring even the
third cumulant with conventional techniques has turned out to be
difficult and so far its only measurements exist for the case of a
tunnel junction. \cite{ReuletBomze} The attention is thus turning
towards using other mesoscopic systems as fluctuation detectors.
\cite{deblock,tobiska,heikkila,pekola,schoelkopf}



In this Paper we analyze the transitions caused by external
fluctuations on a probe quantum system. First, we present a
formula correcting the Golden Rule transition rates by taking into
account the next order effects that are dependent on the third
cumulant.
This is essential in developing generic methods for detecting
non-Gaussian fluctuations. We can establish conditions imposed to
suitable probes 
of third-cumulant induced excitations. Although we concentrate on
current fluctuations, our general analysis is independent of the
physical system as long as the fluctuations are linearly coupled
to the probe system. To demonstrate the results, we consider a
quantum two-state system (qubit) as a probe candidate and 
propose a setup for measuring the effects of the
frequency-dependent third cumulant of current fluctuations by a
Josephson flux qubit.\cite{mooij99,makhlin} This can be viewed as
a generalization of using qubits as spectrometers of the quantum
noise power, \cite{schoelkopf} a method which has already been
experimentally demonstrated.\cite{astafiev}

Our starting point is the Hamiltonian
\begin{align} \label{perus}
H=H_{\mathrm{ext}}+H_{s}+H_{\mathrm{int}},
\end{align}
where $H_{\mathrm{ext}}$ and $H_s$ describe the environment where
the current fluctuates and the quantum system we use as a probe
for the fluctuations, respectively, and $H_{\mathrm{int}}$ is the
interaction Hamiltonian between the environment and the probe.
Motivated by the case of a current-biased Josephson junction and
the magnetic interaction between two circuits considered below, we
study the bilinear coupling of the form
$H_{\mathrm{int}}=g\,\delta I\phi$. Here $\delta I$ is the current
fluctuation operator acting on the environment, $\phi$ is an
operator acting on the probe system and $g$ is the coupling
constant of the interaction. We assume that the quantum system is
described by a set of energy eigenstates $\{|n\rangle
 \}$ and the average current effect $g\,\langle I\rangle\phi$ is included in
 $H_{s}$. 
 Treating $H_{\mathrm{int}}$ as
a perturbation, the Fermi Golden Rule predicts the transition rate
$\Gamma^{(2)}_{n\to n'}=\frac{2\pi\,
g^2}{\hbar^2}|\phi_{n\,n'}|^2S_{\delta
I}(\frac{E_n-E_{n'}}{\hbar})$ between the eigenstates of the probe
system.\cite{schoelkopf} The matrix element is defined as
$\phi_{n\,n'}=\langle n|\phi|n'\rangle$ and the noise power
$S_{\delta
I}(\omega)=\frac{1}{2\pi}\int_{-\infty}^{\infty}e^{i\omega
t}\langle\delta I(t)\delta I(0)\rangle d\,t$. The lowest-order
estimate $\Gamma^{(2)}$ is thus proportional to the second
cumulant of current fluctuations. The correlator in the above
expression is calculated with respect to the environment
Hamiltonian $H_{\mathrm{ext}}$ as if the probe system did not
exist. Below, we correct the transition rate $\Gamma^{(2)}$ by
calculating the next order contribution $\Gamma^{(3)}$, depending
on the third cumulant.

We solve the density matrix evolution using the real-time Keldysh
method, as outlined in Refs.~\onlinecite{schoeller,feynman}, which
is a natural formalism for studying a small subsystem in a larger
environment. We are interested in the dynamics of the probe system
in particular, so we study the reduced density operator
$\rho(t)=\mathrm{Tr}_{\mathrm{ext}}\,\rho_{\mathrm{tot}}(t)$ where
$\rho_{\mathrm{tot}}$ is the density operator for the system and the
environment. The trace goes over a complete set of environment
states. The idea is to solve the temporal evolution of a diagonal
element of the reduced density matrix $\rho_{n'n'}(t)=\langle
n'|\rho(t)|n'\rangle$ with the initial condition
$\rho_{n\,n}(t_0)=1$ $(n\not=n')$. In the long-time limit
$\rho_{n'n'}(t)$ is proportional to the total evolution time
$t-t_0$, the coefficient being the transition rate $\Gamma_{n\to
n'}$. We calculate the rates between well-specified states of the
reduced system. Therefore, without loss of generality, we use an
initial state of the form
$\rho_{\mathrm{tot}}(t_0)=\rho_{\mathrm{ext}}(t_0)\otimes\rho(t_0)$,
where $\rho_{\mathrm{ext}}(t_0)$ describes the initial state of
the environment.

\begin{figure}[h]
\centering
\begin{picture}(100,90)
\put(-70,0){\includegraphics[width=0.99\columnwidth]{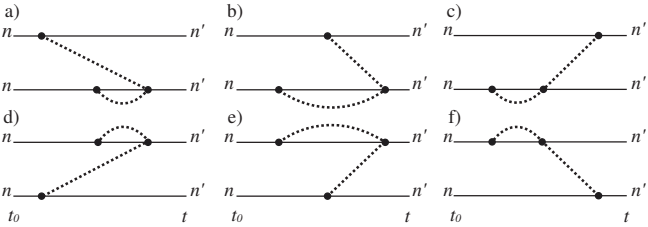}}
\end{picture}
\caption{Six diagrams contributing to $\Gamma_{n\to n'}^{(3)}$. Each
diagram represents temporal evolution from the initial state
$|n\rangle\langle n|$ to the final state $|n'\rangle\langle n'|$.
The dots represent interaction vertices $g\,\delta I\phi$ and the
horizontal lines forward and backward propagation in time.}
\label{keldysh}
\end{figure}
As the lowest-order contribution to 
$\Gamma_{n\to n'}$ is the well-known Golden-Rule result
$\Gamma_{n\to n'}^{(2)}$, we concentrate on the next order
contribution $\Gamma_{n\to n'}^{(3)}$. The total rate is then
given by $\Gamma_{n\to n'}=\Gamma_{n\to n'}^{(2)}+\Gamma_{n\to
n'}^{(3)}$. Treating $H_{\mathrm{int}}=g\,\delta I\phi$ as a
perturbation and using the graphical rules derived in
Ref.~\onlinecite{schoeller}, we find six different diagrams
contributing to $\Gamma_{n\to n'}^{(3)}$, see Fig. \ref{keldysh}.


To present the result in a compact way we define a correlator
\begin{align}\label{kor1}
&\delta^3 I(\omega_1,\omega_2)=
\frac{1}{(2\pi)^2}\int_{-\infty}^{\infty}d(t_3-t_1)\int_{-\infty}^{\infty}d(t_2-t_1)\times\nonumber\\
&\times e^{i\omega_1(t_2-t_1)+i\omega_2(t_3-t_1)} \langle
\widetilde{T} [\delta I(t_1)\delta I(t_2)]\delta I(t_3)\rangle,
\end{align}
where $\widetilde{T}$ denotes the anti-time-ordering operator. The
 time-dependent correlator is calculated with respect to the free
 external Hamiltonian $H_{\mathrm{ext}}$ with the density operator
 $\rho_{\mathrm{ext}}(t_0)$. We
 assume $H_{\mathrm{ext}}$ to be independent of time and
 $\rho_{\mathrm{ext}}(t_0)$ to describe a stationary state with respect
 to $H_{\mathrm{ext}}$. Our results can be also stated with the help of
 the Fourier transform of $\langle
\delta I(t_3)T[\delta I(t_2)\delta I(t_1)]\rangle$, which is the
complex conjugate of the previous correlator, so it is a matter of
choice which one to use. Our definition of the frequency-dependent
third cumulant (\ref{kor1}) differs from the one studied in Ref.
\onlinecite{galaktionov}, which consists of the sum of all
possible Keldysh orderings. Whereas that definition is relevant in
studying the evolution of the off-diagonal density matrix
dynamics, transition rates cannot be obtained from that form. If
the state of the environment is invariant under time reversal as
usually in equilibrium at low magnetic fields, the correlator
(\ref{kor1}) vanishes.

 Evaluating and summing the different contributions shown in Fig. \ref{keldysh},
  we obtain the result
\begin{align}\label{tulos0}
\Gamma^{(3)}_{n\to n'}=\frac{4\pi\,g^3}{\hbar^3}\mathrm{Re}
 \sum_{n_1} \bigg[&\int_{-\infty}^{\infty}\frac{\delta^3I(\frac{E}{\hbar},\frac{E_{n'}-E_n}{\hbar})}{E-(E_{n_1}-E_{n'})-i\eta}dE
 \nonumber\\
 &\times\phi_{n',n}\phi_{n_1,n'}\phi_{n,n_1}].
\end{align}
The summation is extended over all the eigenstates of $H_{s}$ and
$\eta$ denotes a positive infinitesimal quantity. With the help of
the identity $\frac{1}{x-x_0\pm i\eta}=P\frac{1}{x-x_0}\mp
i\pi\delta(x-x_0)$, where $P$ stands for a principle value integral,
we can write (\ref{tulos0}) in the form
\begin{align}\label{tulos}
&\Gamma^{(3)}_{n\to n'}=-\frac{4\pi\,g^3}{\hbar^3}\times
\,\mathrm{Im}
 \sum_{n_1} \left[-i\,P\frac{\delta^3I(\frac{E}{\hbar},\frac{E_{n'}-E_n}{\hbar})}{E-(E_{n_1}-E_{n'})}
 \right.\nonumber\\
 &+\left.\pi
 \delta^3I(\frac{E_{n_1}-E_{n'}}{\hbar},\frac{E_{n'}-E_n}{\hbar})\right]\phi_{n',n}\phi_{n_1,n'}\phi_{n,n_1}.
\end{align}
 Even without the knowledge of $\delta^3
I(\omega_1,\omega_2)$, the general results
(\ref{tulos0},\ref{tulos}) contain some information about the
requirements made for the meter designed to detect the
third-cumulant effects. The structure of the product of the matrix
elements $\phi_{n_in_j}$ restricts the possible physical
realizations used in detecting the transitions induced by the
third cumulant. Generally the operator $\phi$ should either couple
several states of the system, or both matrix elements
$\phi_{n,n}$ and $\phi_{n,n+1}$ should be finite. 


Next we turn to study the case where the probe system is a qubit.
The system Hamiltonian can be written as
$H_{s}=-\frac{1}{2}B_z\sigma_z-\frac{1}{2}B_x\sigma_x$ and the
interaction term as $H_{\mathrm{int}}=g\,\delta I\sigma_z$. 
 The system Hamiltonian has the
eigenstates
$|E_1\rangle=\alpha|\uparrow\rangle+\beta|\downarrow\rangle$,
$|E_0\rangle=-\beta|\uparrow\rangle+\alpha|\downarrow\rangle$ and
the eigenenergies $E_1=\frac{1}{2}\sqrt{B_x^2+B_z^2}$,
$E_0=-\frac{1}{2}\sqrt{B_x^2+B_z^2}$. The coefficients can be
parametrized as $\alpha=\mathrm{cos}\frac{\phi}{2}$ and
$\beta=\mathrm{sin}\frac{\phi}{2}$, where
$\phi=\arctan(\frac{B_x}{B_z})$. 
We denote the energy difference between the two eigenstates as
$\Delta E=\sqrt{B_x^2+B_z^2}$. 
Using the above conventions and the general result (\ref{tulos}),
we can express the corrections to the transition rates as
\begin{align}\label{tulos2}
&\Gamma_{E_1\to E_0}^{(3)}=\frac{16\pi g^3}{\hbar^3}F(\frac{\Delta
E}{\hbar})(\alpha\,\beta)^2(\alpha^2-\beta^2) \nonumber \\
&\Gamma_{E_0\to E_1}^{(3)}=-\frac{16\pi g^3}{\hbar^3}F(-\frac{\Delta
E}{\hbar})(\alpha\,\beta)^2(\alpha^2-\beta^2).
\end{align}
The function $F(\omega)$ contains the information about the third
cumulant and is defined as
\begin{align}\label{funktio}
F(\omega)=\mathrm{Im}\left[-iP\frac{\delta^3I(\frac{E}{\hbar},-\omega)}{E-\hbar
\omega}
+iP\frac{\delta^3I(\frac{E}{\hbar},-\omega)}{E}\right.+\nonumber\\
+\left.\pi\delta^3I(\omega,-\omega)-\pi\delta^3I(0,-\omega)\right].
\end{align}
Comparing the result (\ref{tulos2}) with the Golden Rule rates
\begin{align}\label{golden}
&\Gamma_{E_1\to E_0}^{(2)}=\frac{8\pi g^2}{\hbar^2}S_{\delta
I}(\frac{\Delta
E}{\hbar})(\alpha\,\beta)^2 \nonumber \\
&\Gamma_{E_0\to E_1}^{(2)}=\frac{8\pi g^2}{\hbar^2}S_{\delta
I}(-\frac{\Delta E}{\hbar})(\alpha\,\beta)^2,
\end{align}
one notices that the function $F(\omega)$ plays a similar role in
$\Gamma^{(3)}$ as the noise power in $\Gamma^{(2)}$.

Supposing we can control the effective magnetic fields $B_x$ and
$B_z$, we can optimize the parameters $\alpha$ and $\beta$ to
produce the maximum effect from $\Gamma^{(3)}$. The absolute value
of the expression $(\alpha\,\beta)^2(\alpha^2-\beta^2)$ is
maximized by choosing $\alpha=0.89$ and $\beta=0.46$ or vice
versa, i.e., $B_x=1.4B_z$ or $B_z=1.4 B_x$. 
By changing the magnitude of $\Delta
E=\sqrt{B_x^2+B_z^2}$ but keeping 
$B_x/B_z$ fixed, one can probe $F(\omega)$ as a function of
frequency.

A physical qubit always has some intrinsic noise mechanism, in
solid-state realizations produced by the electromagnetic
environment, which cannot be neglected (we consider the external
fluctuation circuit as an additional environment). To be measurable,
the external current fluctuation effects have to be significant
compared to transitions due to the intrinsic noise.

\begin{figure}[h]
\centering
\begin{picture}(100,90)
\put(-60,0){\includegraphics[width=0.85\columnwidth]{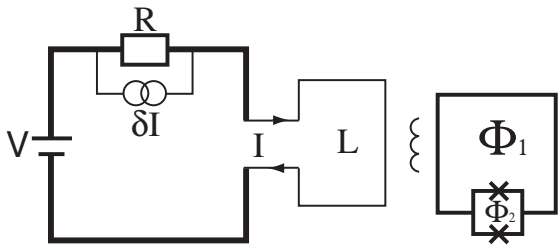}}
\end{picture}
\caption{Josephson flux qubit inductively coupled to an external
circuit producing current fluctuations. The effective magnetic
fields $B_z, B_x$ can be tuned by controlling fluxes
$\Phi_1,\Phi_2$ to maximize the effects of $\Gamma^{(3)}$.}
\label{kohina}
\end{figure}

A possible physical realization for the system considered above is
a Josephson flux qubit \cite{mooij99,makhlin} coupled inductively
to the external circuit, see Fig.~\ref{kohina}. The interaction
Hamiltonian is of the form
$H_{\mathrm{int}}=\frac{M\Delta\phi}{2L_{qb}}\,\delta I\sigma_z$,
where $M$ is the mutual inductance between the qubit and the
external circuit, $\Delta\phi$ is the flux difference between the
two states of the flux qubit and $L_{qb}$ is the inductance of the
qubit. We choose $\langle \delta I \rangle = 0$, since the effects
of the finite average external current can be included in
redefining $B_z$ or eliminated by a flux control. The effective
magnetic fields can be controlled by external fluxes through the
loops so the qubit can be biased to the optimal point for
detecting $\Gamma^{(3)}$. The transition rates follow from
Eqs.~(\ref{tulos2}) and (\ref{golden}) after the identification
$g=\frac{M\Delta\phi}{2L_{qb}}$.  We assume that the energy gap to
the higher states is large compared to any other energy scales in
the system, allowing us to make the two-state approximation and to
neglect the effective interaction terms nonlinear in $\delta I$.


Let us estimate $\Gamma^{(3)}$ in a flux qubit for a specific setup.
Suppose that the external circuit consists of a scatterer with
resistance $R$ and loop inductance $L$. We assume that the third
cumulant of current fluctuations in the scatterer is frequency
independent in the frequency scale of the circuit, $\omega_L\equiv
R/L$. This is generally the case provided that the voltage
 $eV$ over and the Thouless energy $E_T$ of the scatterer, defined as
 the inverse time of flight through it, satisfy $eV,
E_T\gg \hbar \omega_L$.\cite{galaktionov,pilgram} Then the
frequency dependence of the correlator (\ref{kor1}) arises solely
from the classical effect of the inductance $L$ modifying the
noise. In this limit Eq.~(\ref{kor1}) can be approximated by
\begin{equation} \label{estimaatti}
\delta^3I(\omega_1,\omega_2)= \frac{F_3e^2I
(2\pi)^{-1}}{(1+\frac{i\omega_1L}{R})(1+\frac{i\omega_2L}{R})(1-\frac{i(\omega_1+\omega_2)L}{R})},
\end{equation}
where $I$ is the average current in the circuit and $F_3$ is a
  scatterer-specific proportionality constant ("Fano factor") between the
  third cumulant and the current.

In deriving \eqref{estimaatti}, we assumed $L_{qb} I_{qb}^2 \ll L
I^2$ where $I_{qb}$ is the current in the qubit, allowing us to
neglect the back-action of the qubit on these fluctuations. This
leads to the rates $\Gamma^{(3)}_{E_0 \rightarrow
E_1}=\Gamma^{(3)}_{E_1 \rightarrow E_0} \equiv \Gamma^{(3)}$ given
by
\begin{equation}\label{eq:tulos1}
\Gamma^{(3)}= A\frac{\Delta E \omega_L^3}{(\Delta
E^2+\hbar^2\omega_L^2)(\Delta E^2+4\hbar^2\omega_L^2)},
\end{equation}
where $A\equiv32\pi F_3 e^2 I g^3 (\alpha
\beta)^2(\beta^2-\alpha^2)$. The noise power for the setup can be
written as
\begin{equation}
S_{\delta I}(\omega)= \frac{F_2
(eI-\frac{\hbar|\omega|}{R})\theta(eV-\hbar
|\omega|)+\frac{\hbar\omega\theta(\omega)}{R}
}{1+\omega^2/\omega_L^2} \label{freqdepnoise}
\end{equation}
where $F_2$ is the Fano factor for the second cumulant and
$\theta(x)$ is the Heaviside step function. This formula includes
the quantum fluctuations (last term) and is valid for our case
provided that the temperature $T$ is low, $k_B T \ll \Delta E$. In
the limit $eV \gg \Delta E$ we get from Eqs.~(\ref{golden}),
(\ref{eq:tulos1}) and (\ref{freqdepnoise}) that
$\Gamma^{(2)}_{E_1\to E_0}=\Gamma^{(2)}_{E_0\to
E_1}\equiv\Gamma^{(2)}$ and
\begin{align}\label{suhde}
\gamma_3 \equiv
\frac{\Gamma^{(3)}}{\Gamma^{(2)}}=2(\beta^2-\alpha^2)\tilde{g}
\frac{F_3}{F_2} \frac{\Delta E \hbar\omega_L}{\Delta
E^2+4\hbar^2\omega_L^2},
\end{align}
with $\tilde{g}=\left(\frac{M\Delta \phi\, e}{\hbar
L_{qb}}\right)$. For the optimal parameters $\alpha$ and $\beta$
mentioned above, $\alpha^2-\beta^2=0.58$. The phase difference of
the two flux states can be of order $\Phi_0/4=h/8e$, so we may
estimate $\tilde{g} \approx \left(\frac{2\pi\,M}{8
L_{qb}}\right)$. Consequently, it can be made of order unity or
greater by an efficient inductive coupling and a large external
inductance $L$. The factor $F_3/F_2$ depends solely on the nature
of noise produced by the scatterer.\cite{f3note} For realistic
parameters $\omega_L/(2\pi)=10\,\mathrm{GHz}$ and $\Delta
E/h=1\,\mathrm{GHz}$ the last factor is about 2.5\%. Optimizing
the setup one could expect a relative effect
$\left|\frac{\Gamma^{(3)}}{\Gamma^{(2)}}\right|$ up to roughly 10
\%, which shows that the third cumulant effect can be significant.

Now suppose that the intrinsic relaxation of the qubit is caused by
an independent zero-averaged fluctuating Gaussian field. Then the
second-order rate should be replaced by the sum of rates caused by
the field and the external circuit, the third-order rate remaining
unchanged. In the case of a zero-temperature environment, this
intrinsic relaxation rate $\Gamma_{\rm int}$ can be quantified by
the $Q$-factor, $\Gamma_{\rm int}=\Delta E/\hbar Q$. In this case,
its ratio to the rate $\Gamma^{(2)}$ is
\begin{equation}
\gamma_Q^{\rm int}\equiv \frac{\Gamma_{\rm
int}}{\Gamma^{(2)}_{E_0\to E_1}}=\frac{R}{4R_Q} \frac{1}{Q
\tilde{g}^2} \frac{1+\Delta E^2/\hbar\omega_L^2}{F_2 (\frac{eV}{
\Delta E}-1) \theta(eV- \Delta E) (\alpha \beta)^2}.
\end{equation}
Here $R_Q=h/e^2$.

\begin{figure}[h]
\centering
\includegraphics[width=0.9\columnwidth]{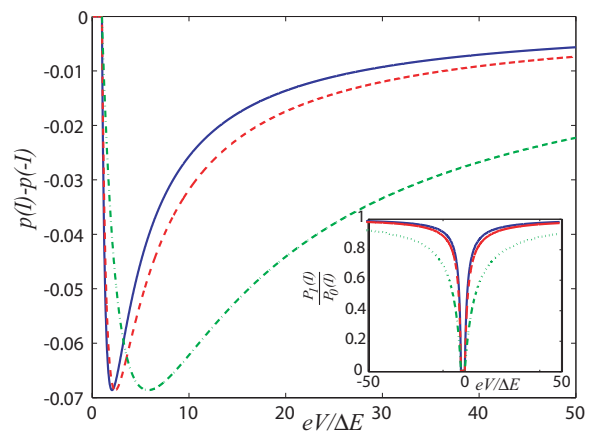}
\caption{(Color online) Difference $p-p'$ in the qubit
polarizations at different directions of the current for different
magnitudes of intrinsic relaxation, $Q=10$ (blue solid line),
$Q=1$ (red dashed line), and $Q=0.1$ (green dash-dotted line).
Inset shows the corresponding total polarization vs. voltage
$V=RI$. Note that the behavior in the region $eV \approx \Delta E$
relies on the approximations made to obtain Eq.~\eqref{suhde}.
These curves have been calculated with $\alpha=0.89$,
$\beta=0.46$, $\tilde{g}=F_2=F_3=1$, $\hbar\omega_L=2 \Delta E$
and $R=0.2 R_Q$.} \label{prob}
\end{figure}

One possibility to detect $\Gamma^{(3)}$ is to let the qubit reach
the stationary state and then determine the probabilities
$P_{E_0}$ and $P_{E_1}=1-P_{E_0}$ of the states $|E_0\rangle$ and
$|E_1\rangle$. This can be achieved by repeated measurements of
the qubit. From detailed balance we get
\begin{equation}
p\equiv \frac{P_{E_1}}{P_{E_0}}
=\frac{\Gamma_{E_0\to E_1}^{(2)}+\Gamma_{E_0\to
E_1}^{(3)}}{\Gamma_{E_1\to E_0}^{(2)}+\Gamma_{E_1\to
E_0}^{(3)}}=\frac{1+\gamma_3}{1+\gamma_Q+\gamma_3},
\end{equation}
where $\gamma_Q= \gamma_Q^{\rm int}+\gamma_Q^{\rm c}$ and
$\gamma_Q^{\rm c}=(\Gamma^{(2)}_{E_1\to E_0}-\Gamma^{(2)}_{E_0\to
E_1})/\Gamma^{(2)}_{E_0\to E_1}$. Now inverting the external
current $I$, $\Gamma^{(3)}$ changes sign and we get $p'\equiv
P'_{E_1}/P'_{E_0}=(1-\gamma_3)/(1+\gamma_Q-\gamma_3)$. From the
above relations one can solve $\Gamma_{E_1\to E_0}^{(3)}$ and
$\Gamma_{E_0\to E_1}^{(3)}$ provided that the probabilities,
$\Gamma_{\rm int}$ and $\Gamma^{(2)}_i$ are known. One can
evaluate $\Gamma^{(2)}_i$ by applying Eq.~(\ref{golden}) or the
rates can be determined experimentally. From Eqs.\! (\ref{tulos2})
and (\ref{golden}) we see that when $\alpha=\beta$, $\Gamma^{(3)}$
vanishes  but $\Gamma^{(2)}_i$ remains finite. By keeping $\Delta
E$ fixed but setting $B_z=0$ it is possible to measure
$\Gamma^{(2)}_i$ independently. Figure \ref{prob} shows the
asymmetry $p-p'$ in the change of polarization with respect to the
current in the source as a function of the magnitude of the
current (bias voltage $V=RI$).

In conclusion, we have studied the transitions induced by the third
cumulant of current fluctuations on a probe quantum system. We have
calculated a general formula for the transition rates and propose a
scheme to measure the predicted results using a Josephson flux
qubit. We have shown that the third-order transition rates are
governed by the variant of the third cumulant which to our knowledge
has not been studied before.

We thank Valentina Brosco, Frank Hekking and Jukka Pekola for
fruitful discussions. TTH acknowledges the funding from the Academy
of Finland.
 
\end{document}